\documentclass{amsart}
\usepackage{amsmath}
\usepackage{amsthm}
\usepackage{amsfonts}
\usepackage{amssymb}
\newtheorem{Theorem}{Theorem}[section]

\newtheorem{Lemma}[Theorem]{Lemma}
\newtheorem{Corollary}[Theorem]{Corollary}
\theoremstyle{definition}

\numberwithin{equation}{section}

\newcommand{\N}{{\mathbb N}}

\begin{document}

\title{Schr\"odinger Operators with Many Bound States}

\author{David Damanik and Christian Remling}

\address{Mathematics 253--37\\
California Institute of Technology\\
Pasadena, CA 91125}

\email{damanik@caltech.edu}

\urladdr{math.caltech.edu/people/damanik.html}

\address{Universit\"at Osnabr\"uck\\
Fachbereich Mathematik\\
49069 Osnabr\"uck\\
Germany}

\email{cremling@mathematik.uni-osnabrueck.de}

\urladdr{www.mathematik.uni-osnabrueck.de/staff/phpages/remlingc.rdf.html}

\date{\today}



\begin{abstract}
Consider the Schr\"odinger operators $H_{\pm}=-d^2/dx^2\pm V(x)$. We present a method for
estimating the potential in terms of the negative eigenvalues of these operators. Among
the applications are inverse Lieb-Thirring inequalities and
several sharp results concerning the spectral properties of
$H_{\pm}$.
\end{abstract}

\maketitle

\section{Introduction}
We are interested in Schr\"odinger equations
\begin{equation}
\label{se}
-y''(x) + V(x)y(x)=Ey(x)
\end{equation}
and the associated self-adjoint operators $H_+=-d^2/dx^2+V(x)$ on $L_2(0,\infty)$.
The potential $V$ is assumed to be locally integrable on $[0,\infty)$.
One also needs a boundary condition of the following form:
\begin{equation}
\label{bc}
y(0)\cos\alpha - y'(0)\sin\alpha = 0 .
\end{equation}
The basic question we would like to address is the following: \textit{What is the
influence of the structure of the discrete spectrum on the potential $V\!$ and on the
spectral properties of $H_+$?}

It is then necessary to consider $H_+$ and $H_-=-d^2/dx^2-V(x)$ simultaneously because
otherwise sign definite potentials provide counterexamples to any possible positive
result one might imagine. So our basic assumption is the following:
\begin{equation}
H_{\pm} \text{ are bounded below and } \sigma_{{\rm ess}}(H_{\pm}) \subset [0,\infty).
\tag{$\Sigma_{{\rm ess}}$}
\end{equation}
Assuming ($\Sigma_{{\rm ess}}$), we can list the negative eigenvalues of $H_+$ and $H_-$
together as $-E_1\le -E_2 \le \ldots$, with $E_n>0$. The list is either finite (or even
empty) or $E_n\to 0$. The $E_n$'s of course depend on the boundary condition \eqref{bc}.
However, we will usually be interested in situations where $\sum E_n^p<\infty$, with
$p\ge 0$, and, by the interlacing property, this condition is independent of $\alpha$.

We deal with the one-dimensional case only in this paper, but our methods are not limited
to this situation. We plan to explore higher-dimensional operators in a future project.

Our original motivation for this work came from the following result, which completely
clarifies the situation when $\{E_n\}$ is a finite set.
\begin{Theorem}[Damanik-Killip \cite{dk} (see also \cite{dks})]
\label{TDK} Assume {\rm (}$\Sigma_{{\rm ess}}${\rm )}. Moreover, assume that $\{ E_n\}$
is finite. Then $\sigma_{{\rm ess}}=[0,\infty)$, and the spectrum is purely absolutely
continuous on $[0,\infty)$ for any boundary condition at $x=0$.
\end{Theorem}
Here the statements refer to $H_+$, say. Of course, since $-V$ satisfies the same
hypotheses as $V\!$, we automatically obtain the same assertions for $H_-$ as well, so
this distinction is actually irrelevant. We keep this convention, however, because it
will help to slightly simplify the formulation of our results.

In \cite{dk}, it is also assumed that $V\in\ell_{\infty}(L_2)$,
that is, $\sup_{x\ge 0} \int_x^{x+1} V^2(t)\, dt <\infty$.
Our treatment below, especially the material from Sect.~3,
will show that this technical assumption is unnecessary.

The aim of this paper is to develop tools for handling arbitrary discrete spectra $\{E_n
\}$, not necessarily finite. At the heart of the matter is a new method for estimating
the potential in terms of the $E_n$'s in general situations. We defer the exact
description of this to Sect.~2 and limit ourselves to a few general remarks in this
introduction (refer to Theorem~\ref{TWQ} below for the full picture).

The most important aspect of our method is this: The rate of convergence with which $E_n$
tends to zero determines the \textit{geometry} of the situation. More precisely, we
obtain intervals $I_n$ whose lengths obey the scaling relation $|I_n| \sim E_n^{-1/2}$.
We will also write $V$ as $W'+W^2$ plus a remainder, with $\|W\|_{L_2(I_n)} \lesssim
|I_n|^{-1/2}$. This representation of $V$ is natural in this context because
$-d^2/dx^2+q(x)$ with Dirichlet boundary conditions ($y=0$) has no negative spectrum if
and only if $q=w'+w^2$ for some $w$. More importantly, it is also very useful in the
applications we want to make in this paper. Note also that any relation between the
potential and the eigenvalues must respect the invariance of the problem under the
rescaling $V(x)\to g^2V(gx)$, $E\to g^2 E$.

To give a more specific impression of what we can do
with our techniques, we mention the following:
\begin{Corollary}
\label{C1.2}
Assume {\rm (}$\Sigma_{{\rm ess}}${\rm )}. Moreover, assume that $\sum E_n^{1/2}<\infty$.
Then there exists $V_0\in L_1(0,\infty)$ so that $H_+ + V_0$ with Dirichlet boundary
conditions has no negative spectrum.
\end{Corollary}
This is indeed immediate from Theorem~\ref{TWQ}, which says that $V-(W'+W^2)\in L_1$ for
a suitable $W$ in this situation. So small eigenvalues can be removed by a small
perturbation. The exponent $1/2$ in the hypothesis is not essential. For instance, it is
also true that if, more generally, $\sum E_n^p<\infty$ with $p\ge 1/2$, then there exists
$V_0\in\ell_{2p}(L_1)$, that is,
\[
\sum_{n=0}^{\infty} \left( \int_n^{n+1} |V_0(x)|\, dx \right)^{2p} < \infty,
\]
so that $H_+ +V_0\ge 0$.

In some instances, our basic problem of obtaining information on the potential from a
knowledge of its eigenvalues can be attacked with completely different tools, called
\textit{sum rules} (aka trace formulae). While this approach is elegant and leads to
very satisfactory results where it works, it is indirect and less systematic and one is
restricted to those combinations of the $E_n$'s that happen to show up in the sum rules
one can produce. See, for instance, \cite{ks,Kup,LNS,NPVY,sz} for recent work on sum
rules.

The converse problem, that is, the problem of estimating the eigenvalues in terms of the
potential, is classical and has received considerable attention over the years. We
mention, in particular, the topic of \textit{Lieb-Thirring inequalities} (see, e.g.,
\cite{LW} for further information on this). For sign-definite potentials, we obtain
\textit{inverse} Lieb-Thirring inequalities as a by-product of our general method; see
Theorem~\ref{TILT} below.

We now turn to discussing the consequences of our method concerning the spectral
properties of $H_{\pm}$. Taking related results into account (see \cite[Theorem~1]{dhks}
and Theorem~\ref{TDK} above), the following does not come as a surprise.
\begin{Theorem}
\label{Tess} Assume {\rm (}$\Sigma_{{\rm ess}}${\rm )}. Then $\sigma_{{\rm
ess}}=[0,\infty)$.
\end{Theorem}
It is now natural to inquire about
the structure of the spectrum on $(0,\infty)$.
\begin{Theorem}
\label{TDe-K}
Assume {\rm (}$\Sigma_{{\rm ess}}${\rm )}. Moreover, assume that $\sum E_n^{1/2}<\infty$.
Then there exists absolutely continuous spectrum essentially supported by $(0,\infty)$.
\end{Theorem}
Killip and Simon have proved earlier the discrete analog of this. The essential
ingredient in their analysis is a sum rule, and they in fact establish the so-called
Szeg\H{o} condition.
See \cite{ks} for these statements, especially Theorems 3 and 7; compare also \cite{sz}.
Under a slightly stronger assumption, we will also prove that on large sets of
energies $E$, the solutions asymptotically look like plane waves. By this we
mean that
\begin{equation}
\label{y-asymp}
y(x)=  e^{i\sqrt{E} x} + o(1) \quad\quad (x\to\infty).
\end{equation}
We let $S$ be the exceptional set where we do \textit{not} have solutions
of this asymptotic form.
So we define
\[
S= \{ E>0: \text{ no solution of \eqref{se} satisfies \eqref{y-asymp}} \} .
\]
Note that if $E\in (0,\infty)\setminus S$, then the complex conjugate of the
solution $y$ from \eqref{y-asymp} is a linearly independent solution of the
same equation, so we have complete control over the solution space for such $E$'s.
In particular, there is no subordinate solution then,
so that the singular part of the spectral measure on $(0,\infty)$
must be supported by $S$ for any boundary condition \eqref{bc}.
\begin{Theorem}
\label{TC-K} Assume {\rm (}$\Sigma_{{\rm ess}}${\rm )}. Moreover, assume that $\sum
E_n^p<\infty$ for some $p<1/2$. Then $|S|=0$.
\end{Theorem}
In particular, this again implies that $H_+$ has absolutely continuous spectrum
essentially supported by $(0,\infty)$. We obtain a more detailed statement here, giving
asymptotic formulae for the solutions, but are unable to treat the borderline case
$p=1/2$. The situation is completely analogous to the known results on operators with
$L_q$ potentials $V\!$. This is no coincidence, because the techniques are the same:
Theorem~\ref{TC-K} crucially depends on work of Christ and Kiselev \cite{CK,CK2}, and the
proof of Theorem~\ref{TDe-K} follows ideas of Deift and Killip \cite{DeK}.

Perhaps somewhat surprisingly, the statement of Theorem~\ref{TC-K} can be sharpened if
$p$ can be taken smaller than $1/4$.
\begin{Theorem}
\label{Tdim} Assume {\rm (}$\Sigma_{{\rm ess}}${\rm )}. Moreover, assume that $\sum E_n^p
< \infty$, with $0\le p< 1/4$. Then $\dim S\le 4p$.
\end{Theorem}
Note that the corresponding statement on $L_q$ potentials is false: There are potentials
$V\in \bigcap_{q>1} L_q$ with $\dim S=1$ (see \cite[Theorem 4.2b)]{Remtams}).

As explained above, Theorem~\ref{Tdim} implies that the singular part of the spectral
measure is supported on a set of dimension $\le 4p$. A related consequence is the fact
that the spectrum is \textit{purely} absolutely continuous on $[0,\infty)$ for all
boundary conditions not from an exceptional set $B\subset [0,\pi)$, where again $\dim
B\le 4p$ (see \cite[Theorem 5.1]{Remdim} for this conclusion).

If, on the other hand, $\sum E_n^p <\infty$ with $1/4\le p < 1/2$, no strengthening of
the statement of Theorem~\ref{TC-K} is obtained, in spite of the stronger hypothesis. The
following theorem shows that no such improvement is possible:
\begin{Theorem}
\label{Tdimopt} Let $e_n>0$ be a non-increasing sequence with $\sum e_n^{1/4}=\infty$.
Then there exists a potential $V\!$ so that {\rm (}$\Sigma_{{\rm ess}}${\rm )} holds,
$E_n\le e_n$, and $\dim S=1$.
\end{Theorem}
Thus the bound from Theorem~\ref{Tdim} is correct at the extreme values $p=0$ and
$p=1/4$. We make the obvious conjecture that it is optimal throughout its range of
validity. The examples used in the proof of Theorem~\ref{Tdimopt} also show that given
$e_n$'s with $\sum e_n^{1/2} = \infty$, there exists a potential so that $E_n\le e_n$ and
$\sigma_{{\rm ac}}=\emptyset$. Hence Theorem~\ref{TDe-K} is optimal, too, and
Theorem~\ref{TC-K} fails to address only the borderline value $p=1/2$. In this context,
the work of Muscalu, Tao, and Thiele \cite{MTT,MTT2} is also relevant.

One of the main difficulties, when estimating $V$ in terms of the $E_n$'s, comes from the
fact that $V$ can take both signs. It is therefore interesting to compare the above
results with the situation for sign-definite potentials, say $V\le 0$. Then only $H_+$
can have negative eigenvalues. In this situation, the control on $V\!$ exerted by the
eigenvalues gets more explicit.
\begin{Theorem}
\label{TILT} Assume {\rm (}$\Sigma_{{\rm ess}}${\rm )}. Moreover, assume that $V\le 0$,
and consider Neumann boundary conditions {\rm (}$\alpha=\pi/2$ in \eqref{bc}{\rm )}.

{\rm a)} For $0<p\le 1/2$, there exists a constant $C_p$ so that
\[
\int_0^{\infty} |V(x)|^{p+1/2}\, dx \le C_p \sum E_n^{p} .
\]

{\rm b)} For $p\ge 1/2$ and $E_0>0$, there exists a constant $C_p(E_0)$ so that
\[
\sum_{n=0}^{\infty} \left( \int_n^{n+1} |V(x)|\, dx \right)^{2p} \le C_p(E_0) \sum E_n^{p} ,
\]
provided that $E_1\le E_0$.
\end{Theorem}
This is reassuring, but we emphasize again that these inequalities do not really catch
the essence of our method. As outlined above, the behavior of $E_n$ governs the geometry
of the situation, and this part of the information gets lost when we pass to global
bounds as in Theorem~\ref{TILT}. This effect is also responsible for the additional
assumption that $\sup E_n \le E_0$ from part b): The intervals $(n,n+1)$ are not adapted
to the underlying geometry. The need for such a restriction is also apparent from the
fact that part b) is not invariant under the rescaling $V(x)\to g^2V(gx)$, $E\to g^2E$.

The estimates from part a) might be called \textit{inverse Lieb-Thirring inequalities.}
Lieb-Thirring inequalities are bounds of the type of part a) but with the opposite sign.
They hold for $p\ge 1/2$. In particular, for $p=1/2$, we have inequalities in both
directions, so $\int |V|$ and $\sum E_n^{1/2}$ are comparable. This is not a new result;
on the contrary, $p=1/2$ is essentially a sum rule, and the best constant is known
(\cite{GGM}, see also \cite{Schm}).

It is clear that we cannot have inverse Lieb-Thirring inequalities for $p>1/2$ because
$V$ can have local singularities so that $V \notin L_q$ for any $q>1$. It is also
important to work with \textit{Neumann} boundary conditions as there are non-zero
potentials $V \le 0$ with no Dirichlet eigenvalues. The whole-line analog of
Theorem~\ref{TILT} also holds and is perhaps more natural for precisely this reason.

As for the spectral properties, Theorem~\ref{TILT} has the following consequences:
\begin{Corollary}
\label{C1.1} Assume {\rm (}$\Sigma_{{\rm ess}}${\rm )}. Moreover, assume that $V\le 0$.

{\rm a)} If $\sum E_n^{1/2}<\infty$, then the spectrum is purely absolutely continuous on
$(0,\infty)$ for all boundary conditions.

{\rm b)} If $\sum E_n<\infty$, then there is absolutely continuous spectrum essentially
supported by $(0,\infty)$.

{\rm c)} If $\sum E_n^p<\infty$ for some $p<1$, then the solutions satisfy WKB-type
asymptotic formulae for Lebesgue almost all energies $E>0$.
\end{Corollary}
Part a) follows because Theorem~\ref{TILT}a) says that $V\in L_1$. As pointed out above,
this part of the corollary has been known before. Rybkin \cite[Theorem~1]{Ryb} has proved
that the assertion of part b) holds if $V \in \ell_2(L_1)$, that is, if
\[
\sum \left( \int_n^{n+1} |V(x)|\, dx \right)^2 < \infty .
\]
This is the ultimate form of a well-known theorem of Deift and Killip \cite{DeK} which
states that there is absolutely continuous spectrum essentially supported by $(0,\infty)$
if $V\in L_1+L_2$. So part b) of the corollary follows from Theorem~\ref{TILT}b).

The asymptotic formula alluded to in part c) reads
\[
y(x,E) = \exp \left( i\sqrt{E} x - \frac{i}{2\sqrt{E}}\int_0^x V(t)\, dt \right) + o(1)\quad\quad
(x\to\infty) .
\]
Christ and Kiselev \cite{CK} prove that this holds at almost all energies if
$V\in\ell_p(L_1)$ for some $p<2$, so Theorem~\ref{TILT}b) also implies part c) of the
corollary.

As above, these results are complemented by the following:
\begin{Theorem}
\label{TL1opt} Let $e_n>0$ be a non-increasing sequence with $\sum e_n=\infty$. Then
there exists a potential $V\le 0$ so that {\rm (}$\Sigma_{{\rm ess}}${\rm )} holds,
$E_n\le e_n$, and $\sigma_{{\rm ac}}=\emptyset$.
\end{Theorem}

We organize this paper in the obvious way: Section~2 gives a detailed discussion of our
general method. The subsequent sections are concerned with the applications of this to
the spectral theory of $H_{\pm}$, in the order suggested by this introduction. In the
final section, we present the examples announced in Theorems~\ref{Tdimopt} and
\ref{TL1opt}.

\medskip

\noindent\textit{Acknowledgments.} It is a pleasure to thank Rowan Killip and Barry Simon
for useful conversations. C.\ R.\ would like to express his gratitude for the hospitality
of Caltech, where this work was begun.

\section{A Method for Estimating $V$}
As in the previous section, let $H_{\pm}=-d^2/dx^2 \pm V(x)$. We write $H_{\sigma}$ if we
work with one of these operators, but do not want to specify which one. Boundary
conditions (where necessary) will \textit{always} be Dirichlet boundary conditions
($y=0$).

The following theorem may be viewed as the principal result of this paper. Things become
slightly easier in the whole-line setting because we avoid the somewhat artificial
technical problems associated with the effect that a boundary condition can screen part
of the potential. So we discuss this case first. The modifications needed to handle
half-line problems will be described after having completed the treatment of the
whole-line case.
\begin{Theorem}
\label{TWQ} Consider $H_{\pm}$ on $L_2(\mathbb R)$. Assume {\rm (}$\Sigma_{{\rm
ess}}${\rm )}. Then there exist a partition of $\mathbb R$ into intervals $J_n^{(k)}$
with disjoint interiors and a decomposition $V=W'+Q$ with the following properties:

{\rm a)} {\rm (basic properties of $W,Q$)} $W$ is absolutely continuous. If $\sum
E_n^{1/2}<\infty$, then $W\in L_2(\mathbb R)$ and $Q\in L_1(\mathbb R)$.

{\rm b)} {\rm(geometry of the intervals)} The indices $k,n$ vary over the following sets:
$n\in\mathbb N$ and $k\in\mathbb Z$, $-N_n-1<k<N'_n+1$, where $N_n,N_n'\in\N_0\cup\{
\infty \}$. We choose the natural numbering with respect to $k$, that is, $J_n^{(k)}$
lies to the left of $J_n^{(k+1)}$. Then, if we denote the length of $J_n^{(0)}$ by
$\ell_n$, we have that
\[
2^{|k|-4} \ell_n \le \left|J_n^{(k)}\right| \le 2^{|k|-2} \ell_n.
\]
In particular, we can have $N_n=\infty$
for at most one index $n$ and also $N'_{n}=\infty$ for at most one $n$.

{\rm c)} {\rm (detailed estimates on $W,Q$)} If $J=J_n^{(k)}$ for some $n,k$, then
\[
\int_J W^2(x)\, dx \le \frac{10^3}{|J|},\quad
\int_J |Q(x)|\, dx \le \frac{10^3}{|J|} .
\]

{\rm d)} {\rm (growth of the lengths)} $\ell_n \ge 4 E_n^{-1/2}$.
\end{Theorem}
We give explicit constants in these estimates;
this will avoid any misgivings one might possibly have about hidden dependencies
of our constants. However, no attempt will be made to optimize these constants
because their values do not matter to us here.

We start with some preparations. Lemmas~\ref{LWQ} and \ref{Lremove} will be the basic
ingredients to the proof of Theorem~\ref{TWQ}.
\begin{Lemma}
\label{LWQ} Suppose that $H_{\pm}\ge -\epsilon$ on $L_2(a,b)$ for some $\epsilon\ge 0$.
Then, on $(a,b)$, we can write $V=W'+Q$ where $W,Q$ satisfy the following estimates:
\[
\int_a^b (\varphi(x)W(x))^2 \, dx \le \epsilon \int_a^b
\varphi^2(x)\,dx + \int_a^b \varphi'^2(x)\, dx
\]
for all $\varphi\in H_1(a,b)$ with $\varphi(a)=\varphi(b)=0$. Moreover,
\[
|Q(x)| \le 2 \left( \epsilon^{1/2} + \frac{1}{\text{\rm dist}(x, (a,b)^c)} \right) |W(x)|
\]
for all $x\in (a,b)$.
\end{Lemma}
\begin{proof} The hypothesis that $H_{\pm}\ge -\epsilon$ implies that there are zero-free (on
$(a,b)$) solutions $u,v$ of $-u''+Vu=-\epsilon u$ and $-v''-Vv=-\epsilon v$,
respectively. (We really have to take the open interval; if $\epsilon>0$ has been taken
as small as possible, at least one of the functions $u,v$ has zeros at both endpoints.
This explains why the estimates on $Q$ get worse as we approach the endpoints.)

As in \cite{dk}, define
\begin{equation}
\label{1.2}
\gamma = \frac{1}{2} \left( \frac{u'}{u}+\frac{v'}{v}\right) ,\quad
W = \frac{1}{2} \left( \frac{u'}{u}-\frac{v'}{v}\right) .
\end{equation}
Then
\begin{gather}
\label{1.1}
\gamma'  = -\gamma^2 + \epsilon - W^2 \\
V = W'+2\gamma W \equiv W' + Q\nonumber
\end{gather}
\eqref{1.1} is the Schr\"odinger equation, with potential $\epsilon-W^2$ and zero energy, in Riccati form.
Since \eqref{1.2} explicitly displays a solution $\gamma$ on
all of $(a,b)$ and since $\gamma=y'/y$ (or rather $y=\exp (\int^x \gamma)$)
transforms back to the Schr\"odinger form
of \eqref{1.1}, it is clear that $-d^2/dx^2+\epsilon-W^2\ge 0$ on $(a,b)$. The estimate on $W$
now follows at once by using $\varphi$ as a test function in the quadratic form of this operator.

To analyze $\gamma$, we compare with solutions $\gamma_0$ of $\gamma'_0=\epsilon-\gamma_0^2$. We
begin with the case when $\epsilon>0$.
The following holds: If $\gamma(x)=\gamma_0(x)$ for some $x\in (a,b)$,
then $\gamma_0(t)\le \gamma(t)$ for all $t\le x$ from this interval and $\gamma_0(t)\ge\gamma(t)$
if $t\ge x$. Now fix $x\in (a,b)$. Then, by what has just been noted, the value of $\gamma$ at this
$x$ must have the property that the solution $\gamma_0$ with $\gamma_0(x)=\gamma(x)$ satisfies
$\sup_{a+\delta\le t\le x} \gamma_0(t) <\infty$ and $\inf_{x\le t\le b-\delta} \gamma_0(t)>-\infty$ for
all $\delta>0$. It follows from these conditions that
\[
-\epsilon^{1/2} \coth\epsilon^{1/2}(b-x) \le \gamma(x) \le \epsilon^{1/2}\coth\epsilon^{1/2}(x-a) .
\]
This can be seen by again writing the equation for $\gamma_0$ in Schr\"odinger form ($\gamma_0=y'/y$,
$y''=\epsilon y$) and by noting that the largest (smallest) solution $\gamma_0$ comes from a $y$
that has a zero at $a$ (at $b$). Since $\coth x \le 1+x^{-1}$ for $x>0$, we see that
\begin{equation}
\label{gammaaux}
|\gamma(x)| \le \epsilon^{1/2} + \frac{1}{\text{\rm dist}(x, (a,b)^c)} .
\end{equation}
This bound in fact works for all cases ($\epsilon\ge 0$). One can either take suitable
limits or make slight adjustments in the above arguments. As $Q=2\gamma W\!$, the
asserted estimate on $Q$ is an immediate consequence of \eqref{gammaaux}.
\end{proof}
Next, we show that the eigenvalue $-\epsilon$ can already be seen on a length scale
$L\sim \epsilon^{-1/2}$. We need the following calculation with quadratic forms (which
appears to go back to Jacobi \cite{jac}; see also Courant-Hilbert \cite[p.~458]{ch}):
\begin{Lemma}
\label{L1.1} Suppose $-f''+Vf=Ef$, $\varphi\in H_1(a,b)$,
$(\varphi^2 ff')(a)=(\varphi^2 ff')(b)$. Then
\[
\int_a^b \left[ (\varphi f)'^2 + V(\varphi f)^2\right] = \int_a^b
\varphi'^2 f^2 + E\int_a^b \varphi^2 f^2 .
\]
\end{Lemma}
\begin{proof}
An integration by parts shows that
\begin{align*}
\int_a^b \varphi^2 f'^2 & = \varphi^2 ff' \Bigr|_a^b -
\int_a^b f\left( \varphi^2f' \right)'
= -2\int_a^b \varphi\varphi'ff' - \int_a^b \varphi^2ff''\\
& = -2\int_a^b \varphi\varphi'ff' + \int_a^b \varphi^2(E-V)f^2 .
\end{align*}
Plug this into
\[
\int_a^b (\varphi f)'^2 = \int_a^b \varphi'^2f^2 + 2\int_a^b \varphi\varphi'ff'
+\int_a^b \varphi^2f'^2
\]
to obtain the lemma.
\end{proof}
\begin{Lemma}
\label{Lremove} Assume that the smallest eigenvalue of $H_{\sigma}$ on $(a,b)$ {\rm
(}call it $-\epsilon${\rm )} is negative. If $b-a\ge 6\epsilon^{-1/2}$, then there exists
a subinterval $I \subset (a,b)$ of length $|I|=6\epsilon^{-1/2}$ so that $H_{\sigma}$ on
$I$ has an eigenvalue $\le -\epsilon/2$.
\end{Lemma}
In contrast to the previous two lemmas, we now allow unbounded intervals $(a,b)$ as well.
\begin{proof}
Let $f$ be the corresponding eigenfunction,
so $-f''+\sigma Vf=-\epsilon f$ and $f=0$ at the finite endpoints of $(a,b)$.
Let $L=\epsilon^{-1/2}$, and pick a $c$ that
maximizes $\int_{c-L}^{c+L} f^2$. Define
\[
\varphi(x) = \begin{cases} 1 & |x-c|\le L, \\
3/2-|x-c|/(2L) & L<|x-c|<3L, \\
0 & |x-c|\ge 3L. \end{cases}
\]
We now take $I$ as the support of $\varphi$, intersected with $(a,b)$. It can of course
happen that $|I|<6\epsilon^{-1/2}$, but it certainly suffices to show that $H_{\sigma}$
on $I$ has an eigenvalue $\le -\epsilon/2$. We can then simply replace $I$ by a larger
interval $I'\supset I$ of the desired length. By the min-max principle, the eigenvalues
will only go down.

The set $\{ x\in I: \varphi'(x)\not=0 \}$ consists of at most two intervals of length
$\le 2L$ each, so by the choice of $c$, we have that $\int_I \varphi'^2 f^2 \le (1/2L^2)
\int_I \varphi^2f^2$. The function $\varphi f$ is in the form domain of the operator
$H_{\sigma}$ on $I$ (call this form $Q_I$), and Lemma~\ref{L1.1} shows that
\[
Q_I(\varphi f) = \int_I \varphi'^2 f^2 -\epsilon \int_I \varphi^2 f^2
\le \left( \frac{1}{2L^2}-\epsilon \right) \int_I \varphi^2f^2 = -\frac{\epsilon}{2}
\int_I \varphi^2f^2.
\]
Hence $H_{\sigma}$ on $I$ indeed has an eigenvalue $\le -\epsilon/2$.
\end{proof}
We are now ready for the proof of Theorem~\ref{TWQ}. It is probably advisable not to pay
too much attention to the specific evaluation of the constants in the first reading of
the following proof. In most cases, it is almost immediate that an inequality of the type
$a\le Cb$ holds while finding a concrete value for such a $C$ usually requires an
additional (but elementary) calculation.
\begin{proof}[Proof of Theorem~\ref{TWQ}.]
We will present a method for inductively finding intervals with the required properties.
In the main part of this proof, the symbols $I_n$, $J_k$ will be used for these intervals
and we will write $L_n=|I_n|$ for their lengths. We avoid using the letters from the
statement of the theorem from the beginning, because our initial choices will be modified
later on. However, it is true that the $J_n^{(0)}$'s from the theorem are essentially the
$I_n$'s, and the $J_n^{(k)}$'s for $k\not= 0$ correspond to suitable $J_k$'s. Here is an
outline of the method. Put $\epsilon_1=E_1$. The basic idea is to use Lemma~\ref{LWQ} to
write $V=W_1'+Q_1$ with $W_1$, $Q_1$ satisfying certain inequalities. Then we remove an
interval $I_1$, $|I_1|=6\epsilon_1^{-1/2}$ with the properties stated in
Lemma~\ref{Lremove}. On $I_1$, we keep the $W_1$, $Q_1$ just constructed. In the next
step, we consider $H_{\pm}$ on $S_2=\mathbb R\setminus I_1$. We update $\epsilon$ and
obtain a new value $\epsilon_2$. Note that $\epsilon_2\le\epsilon_1$; in fact, after some
steps the improvement must be substantial, because the min-max principle says that if
$N(E)$ denotes the number of eigenvalues below $-E$ of the original operators, then there
can be at most $N(E)$ disjoint intervals with ground state energies $\le -E$. On $S_2$,
we construct a new $V=W_2'+Q_2$ decomposition with the help of Lemma~\ref{LWQ}. The idea
is that since $\epsilon_2\le\epsilon_1$, these new functions admit improved bounds. We
again remove an interval of length $|I_2|=6\epsilon_2^{-1/2}$ by using
Lemma~\ref{Lremove}, and we continue on $S_3=S_2\setminus I_2$. On $I_2$, the functions
$W_2$, $Q_2$ are our final choices. In reality, the procedure does not work quite as
smoothly because the estimates coming from Lemma \ref{LWQ} get worse if $I_n$ lies close
to the boundaries of the current $S_n$.

We now give the details. Put $S_1=\mathbb R$ and write the
smallest eigenvalue of $H_{\pm}$ on $S_1$ as $-\epsilon_1$ ($\epsilon_1>0$). This first step
is easier than the general step because most of the technical problems come from the influence
of the complement of $S_n$, and $S_1^c=\emptyset$.

Apply Lemma~\ref{Lremove} to obtain an interval $I_1\subset S_1$ of length $L_1\equiv
6\epsilon_1^{-1/2}$. $H_{\sigma}$ for suitable $\sigma=\pm$ has an eigenvalue $\le
-\epsilon_1/2$ on $I_1$. We also use Lemma~\ref{LWQ} to write $V$ as $V=W_1'+Q_1$. Define
a test function $\varphi$ as follows: $\varphi\equiv 1$ on $3I_1$, $\varphi\equiv 0$
outside $5I_1$, and $\varphi$ is linear on the remaining two intervals. Here, we denote
by $kI$ the interval of length $k|I|$ with the same center as $I$. By using this
$\varphi$, we obtain from Lemma~\ref{LWQ} that
\[
\int_{3I_1} W_1^2 \le \int \varphi^2 W_1^2 \le \epsilon_1 \int \varphi^2 +\int\varphi'^2 .
\]
Evaluate the integrals and recall that $\epsilon_1=36/L_1^2$. This gives the bound
\begin{equation}
\label{1.5}
\int_{3I_1} W_1^2 \le \frac{134}{L_1} .
\end{equation}
To estimate $Q$, we again use Lemma~\ref{LWQ}. Observe that the second term in
parentheses from the bound on $Q$ is $\le L_1^{-1}$ on $3I_1$ if we take $(a,b)=5I_1$.
Hence the Cauchy-Schwarz inequality together with the bound on $\|W\|_{L_2(3I_1)}$
already proved show that
\begin{equation}
\label{1.10}
\int_{3I_1} |Q_1| \le \frac{281}{L_1} .
\end{equation}
We put $S_2=\mathbb R\setminus I_1$; this concludes the first step. Let us summarize what
we have achieved: First, $H_+$ or $H_-$ has an eigenvalue $\le -\epsilon_1/2$ on $I_1$.
Second, we also have defined $W,Q$ on $I_1$ and estimated these functions there. In fact,
we have done this on the larger interval $3I_1$; this additional information will be
useful later.

We now move on to the general step. The situation is as follows: Our set $S_n$ is a
collection of finitely many disjoint intervals, up to two of them being half lines
(usually there will be exactly two half lines). For a bounded component $(a,b)\subset
S_n$, there are intervals immediately to the left and right, respectively, of $(a,b)$
that were generated in earlier steps. Call them $I_-$ and $I_+$, respectively. By what
has just been observed, they are also equal to $I_j$, $I_k$ for suitable indices $j,k<n$.
The construction we are about to describe makes sure that $b-a\ge 2L_-, 2L_+$, where the
$\pm$ notation is used with the same meaning as above, that is, $L_-=L_j$ and $L_+=L_k$.
Moreover, in such a situation, we have $W'+Q$ decompositions with control of the type
\eqref{1.5}, \eqref{1.10} on $3I_{\pm}\cap (a,b)$. This follows from the fact that these
intervals were generated in previous steps.

To run step $n$, first update $\epsilon$. More specifically, write $-\epsilon_n$ for the
smallest eigenvalue of the operators $H_{\pm}$ on $S_n$. Then, by the min-max principle,
$S_n\subset S_{n-1}$ implies that $\epsilon_n\le\epsilon_{n-1}$. Choose a component
$(a,b)$ of $S_n$ so that the eigenvalue $-\epsilon_n$ occurs there. We will assume that
$b-a<\infty$; in the half-line case, the discussion is similar but easier. Put
$L_n=6\epsilon_n^{-1/2}$ and use Lemma~\ref{LWQ} with $\epsilon=\epsilon_n$ to define
$W_n$, $Q_n$ on $(a,b)$. Several cases arise:

a) \textit{No conflict with the boundary:} This is the easy case because the machine
based on Lemmas~\ref{LWQ} and \ref{Lremove} works smoothly. The precise condition we will
use is that $7\widetilde{I}_n\subset (a,b)$, with $\widetilde{I}_n$ chosen according to
Lemma~\ref{Lremove}.

Let $I_n=\widetilde{I}_n$. Lemma~\ref{LWQ} with a tent-shaped test function supported on
$5I_n$ shows that we have the following analogs of \eqref{1.5}, \eqref{1.10}:
\begin{equation}
\label{estWQ}
\int_{3I_n}W_n^2 \le \frac{134}{L_n},\quad \int_{3I_n} |Q_n| \le \frac{281}{L_n} .
\end{equation}
Let $S_{n+1}=S_n\setminus I_n$ and proceed with step $n+1$. Note that the two newly generated intervals
$(a,b)\setminus I_n$ satisfy the condition that their length is at least twice the length of their neighbors.
This obviously holds if this neighbor is taken to be $I_n$,
and it also holds for their other neighbors because
these intervals come from earlier steps and $L_j\le L_n$ if $j\le n$. This inequality in
turn follows from the definition $L_i=6\epsilon_i^{-1/2}$
and the fact that $\epsilon_i$ is non-increasing.

b) \textit{The extreme opposite of case a):} Assume now that $b-a< L_n$, so that applying
Lemma~\ref{Lremove} is out of the question. We will just remove the whole interval
$(a,b)$ and obtain control on $W_n$, $Q_n$ by what we call the \textit{boundary method.}
To (slightly) simplify the notation, we relabel $a\to 0$, $b\to 2L$ for the time being.
Then $L_{\pm}\le L$. We also drop the index $n$ for the functions $W_n$, $Q_n$.
\begin{Lemma}[The boundary method]
\label{Lbm} In the situation described above, determine $L_0\in (L_-/4,L_-/2]$ so that
$2^N L_0=L$ for some $N\in \mathbb N$. Define $J_k = [2^{k-1}L_0, 2^k L_0]$ {\rm
(}$k=1,\ldots , N${\rm )}. Then
\begin{equation}
\label{estWQbm}
\int_{J_k} W^2 \le \frac{6}{|J_k|}, \quad \int_{J_k} |Q| \le \frac{13}{|J_k|} .
\end{equation}
\end{Lemma}
\begin{proof}[Proof of Lemma~\ref{Lbm}.]
This follows as above (compare \eqref{estWQ}) from a straightforward application of
Lemma~\ref{LWQ} with test functions of tent form supported by $3J_k$ and equal to $1$ on
$J_k$.
\end{proof}
The estimates \eqref{estWQbm} are worse than \eqref{estWQ} in that we need to cut the
interval $[0,L]$ into smaller pieces. The reason for this is that we are close to
possibly large eigenvalues (on $I_-$).

We can now apply Lemma~\ref{Lbm} and the analog of this from the right to get estimates
on $W_n$, $Q_n$ on $(a+L_-/2, b-L_+/2)$. An additional issue needs to be addressed: We
must match $W_n$ with the $W$'s coming from $I_{\pm}$.
\begin{Lemma}
\label{Lmatching} Suppose that $V=w'+q$ on $[c,d]$. It is then possible to define new
functions $W,Q$ so that we still have $V=W'+Q$, but also $W(c)=0$. Moreover, the
following estimates can be achieved:
\[
|W(x)|\le |w(x)|\quad (c\le x \le d), \quad \int_c^d |Q(x)|\, dx \le \int_c^d |q(x)|\, dx + 2|w(c)| .
\]
\end{Lemma}
\begin{proof}[Proof of Lemma~\ref{Lmatching}.]
Define, for $t>0$ (and typically small),
\[
\chi(x)= \begin{cases} 1 & x > c+t, \\
(1/t)(x-c) & c\le x\le c+t, \end{cases}
\]
and let $W=\chi w$, so $Q=q + ((1-\chi)w)'$. Clearly, $|W|\le |w|$, and
\[
\int_c^d |Q| -\int_c^d |q| \le \int_c^{c+t} |w'| + \int_c^d |\chi' w|
\le \int_c^{c+t} \left( |V|+|q| \right) +
(1/t) \int_c^{c+t} |w| \to |w(c)|
\]
as $t\to 0+$. So if $w(c)\not=0$, a sufficiently small $t>0$ will give the desired estimates.
\end{proof}
The lemma is useful in our situation because a weak-type estimate shows that $W$ is small
most of the time. More precisely, consider the interval $J=(L_-/2,L_-)$. Then, writing
$W_-$ for the $W$ obtained in the step in which the interval $I_-$ was removed, it
follows from \eqref{estWQ} that
\begin{equation}
\label{1.6}
\left| \{ x\in J: |W_-(x)|> CL_-^{-1} \} \right| \le \frac{134}{C^2}\, L_- .
\end{equation}
Since $L\ge 2L_-$, this estimate also holds with $W_n$ in place of $W_-$. In particular,
since $134/24^2<1/4$, there exists an $x_0\in J$ so that $|W_-(x_0)|, |W_n(x_0)|\le 24
L_-^{-1}$. We can now apply Lemma~\ref{Lmatching} (and its mirror version to the left)
with $c=x_0$ to obtain a modification of $W_-$, $W_n$, $Q_-$, $Q_n$ in a neighborhood of
$x_0$ with both $W$'s now vanishing at $x_0$. In particular, we can now change from $W_-$
to $W_n$ at $x_0$ without destroying the absolute continuity of $W\!$.

Since $L_0\le x_0< 4L_0$, we have that $x_0\in J_k$ either for $k=1$ or for $k=2$. Fix this $k$. Then
\[
\int_{J_k} |Q| \le \frac{13}{|J_k|} + \frac{281}{L_-} + \frac{4\cdot 24}{L_-} .
\]
Indeed, the first two terms on the right-hand side are the old bounds on $\int |Q_n|$ and
$\int |Q_-|$, respectively, while the last term collects the contributions $2|W_n(x_0)|$,
$2|W_-(x_0)|$, which come from Lemma~\ref{Lmatching}. Since $L_-\ge 2L_0\ge |J_k|$, we
finally obtain $\int_{J_k} |Q| \le 390/|J_k|$. Similarly, $\int_{J_k} W^2 \le 140/|J_k|$.
We do not want to bother about whether or not different $W$'s have been matched on a
given interval, so we simply change the constants for all $k$ to these new values.

So far, we have treated the left half of $(a,b)$ (temporarily denoted by $(0,2L)$ for convenience).
Now apply the mirror version of this to the right half of $(a,b)$.

We summarize: We have basically subdivided $(a,b)$ into two series of intervals $J_k$
(one coming from the left, the other from the right) with geometrically increasing
lengths $\approx 2^k L_{\pm}$. We qualify this by saying ``basically'' because we also
modify the neighboring intervals $I_{\pm}$: We add a piece of length $L_0$, where $L_0$
has the same meaning as above (and of course depends on whether we are considering $I_-$
or $I_+$). This is exactly that part of $(a,b)$ that has not been covered by the $J_k$'s.
For later use, we record this modification and call the enlarged intervals
$I_{\pm}^{(1)}$. Note, however, that the original intervals might also be used again.
More specifically, this happens if the boundary method is also applied to the left of
$I_-$ (or to the right of $I_+$) at a later stage. If, conversely, this has happened
before step $n$, then we already have modifications of these intervals and we then denote
the new modification by $I_{\pm}^{(2)}$. (In fact, these conventions are a bit pedantic
and we could also discard the original intervals right away except that then the
condition that components of $S_{n+1}$ are at least twice the size of the neighboring
$I_k$'s may be violated.)

We have extended the $V=W'+Q$ representation to the $J_k$'s, with the following estimates on $W$
and $Q$:
\begin{equation}
\label{1.11} \int_{J_k} W^2(x)\, dx \le \frac{140}{|J_k|}, \quad \int_{J_k} |Q(x)|\, dx
\le \frac{390}{|J_k|}.
\end{equation}
We put $S_{n+1}=S_n\setminus [a,b]$ and proceed with step $n+1$.

c) \textit{$\widetilde{I}_n$ close to precisely one boundary point:} This case arises
when $a\in 7\widetilde{I}_n$, $b\notin 7\widetilde{I}_n$ or conversely. Let us assume the
first situation. Also recall that $\widetilde{I}_n$, as always in this proof, is a
subinterval of $[a,b]$ of length $L_n=6\epsilon_n^{-1/2}$ chosen according to
Lemma~\ref{Lremove}. We let $I_n=L_n+\widetilde{I}_n$, that is, we take the copy of
$\widetilde{I}_n$ immediately to the right of the original choice. We apply the boundary
method (Lemma~\ref{Lbm}) to obtain control of the type \eqref{estWQbm} on $[a+L_0,\inf
I_n]$, where $L_-/4<L_0\le L_-/2$. In other words, we cover this interval, which has a
length between $L_n/2$ and $4L_n$, by a series of intervals $J_k$ of geometrically
increasing lengths $|J_k|=2^{k-1}L_0$. Recall also in this context that $L_-\le L_n$.
Next, we again match $W_-$ and $W_n$ on $(a+L_-/2,a+L_-)$ with the help of
Lemma~\ref{Lmatching}. As explained above, this leads to the estimates \eqref{1.11}.
Finally, in analogy to case b), we also modify the left neighbor $I_-$ by adding
$[a,a+L_0]$ and call the new interval $I_-^{(k)}$ if this was the $k$th modification (so
$k=1$ or $k=2$).

On the remaining part of $3I_n$, we use Lemma~\ref{LWQ} with a suitable tent-shaped test
function $\varphi$. More precisely, $\varphi=1$ on this set, that is, on the right
two-thirds of $3I_n$. Moreover, $\varphi=0$ outside the interval of size $4L_n$ that is
centered at the right endpoint of $I_n$. We thus see that
\begin{equation}
\label{1.12}
\int W_n^2(x)\, dx \le \frac{98}{L_n},\quad \int |Q_n(x)|\, dx \le \frac{196}{L_n} ,
\end{equation}
where the integrations are over $I_n$ and the interval to the right of $I_n$
of the same size (in other words,
over the part of $3I_n$ not already treated by the boundary method). The constants are smaller
here because this set is also smaller than $3I_n$. However, we will not insist on this; rather, we
use the values from \eqref{estWQ} in this case as well.

We delete $I_n$ and everything to the left of $I_n$ (inside $[a,b]$, that is) to obtain our new set $S_{n+1}$.
The final choice of $W,Q$ has also been described above ($W_n$, $Q_n$ except on an initial
piece of size $\approx L_-$). Note that $S_{n+1}$ now has the same number of components
as $S_n$. The interval $(a,b)$ was replaced by $(a_1,b)$ with $a_1>a$. However, since
$b\notin 7\widetilde{I_n}$, the new component $(a_1,b)$ still satisfies our condition that its length
is at least twice the
lengths corresponding to the neighboring intervals. For the right neighbor, this again follows from the
fact that this right neighbor equals $I_j$ for a suitable $j<n$ and $L_j\le L_n$.

d) \textit{$\widetilde{I}_n$ close to both boundary points:} In other words, $a,b\in
7\widetilde{I}_n$, but, in contrast to case b), $L_n \le b-a$. This case does not require
new ideas; indeed, d) and b) could have been subsumed under one case. Use the boundary
method from both $a$ and $b$, and match $W_n$ with $W_-$ and $W_+$. We obtain two series
of intervals $J_k$ and the estimates from \eqref{1.11} on $W,Q$ on the intervals $J_k$.
We define, as usual, $S_{n+1}=S_n\setminus [a,b]$. So we have in fact just removed the
component $(a,b)$.

All cases have been covered now and we can run the algorithm. One final adjustment is
necessary, however. It may happen that $S_n$ for some $n$ contains components $(a,b)$
with $H_{\pm}\ge 0$ there. As long as at least one of $H_{\pm}$ has negative eigenvalues
on $S_n$, such an interval will not be dealt with by the algorithm. If $b-a<\infty$, we
treat these intervals with the method of case b) as soon as they arise. This is in fact
the obvious thing to do because formally $L=6\epsilon^{-1/2}=\infty$ (since $\epsilon=0$
by assumption). So the boundary method gives us two sequences of intervals $J_k$ with the
usual properties (see Lemma~\ref{Lbm} and the discussion of case b)). We then continue
the algorithm with a new $S_n$ from which these components $(a,b)$ have been removed.

Similarly, if $H_{\pm}\ge 0$ on a half line $(a,b)$, the boundary method produces an infinite
series of intervals $J_k$ that cover $(a,b)$. The usual estimates on $W$ and $Q$ hold, and
$|J_k|\approx 2^{|k|}L$, where $L$ is the length of the neighbor of $(a,b)$. We again remove
$(a,b)$ from $S_n$ and continue.

We can now be sure that our inductive construction produces intervals that cover all of
$\mathbb R$. Indeed, intervals $(a,b)$ on which $H_{\pm}\ge 0$ are treated immediately,
and if $H_{\sigma}$ has a negative eigenvalue $-\epsilon$ on $(a,b)$, the min-max
principle guarantees that there are only finitely many other intervals with eigenvalues
$<-\epsilon$, and after the corresponding number of steps, at the latest, the algorithm
will take care of $(a,b)$. It may of course happen that only part of $(a,b)$ is removed
then, but the length of this removed part is at least $\min\{ L_n, b-a \}$, and
$L_n=6\epsilon_n^{-1/2}$ increases, so a finite number of such steps will suffice to
cover all of $(a,b)$.

To finally verify the assertions of Theorem~\ref{TWQ}, we first need to relabel our
intervals. The $J_n^{(0)}$'s are essentially the intervals $I_n$ that arise in cases a)
and c), but with possible later modifications taken into account. Recall that these
modifications occur if the boundary method is applied at a later stage with $I_n$ taking
the role of a left or right neighbor of the interval currently under consideration. In
other words, we let $J_n^{(0)}=I_n^{(j)}$, with $j$ (which counts the number of
modifications) maximal. Recall also that each of these modifications consists of adding
an interval of length $L$, with $L_n/4<L\le L_n/2$, to $I_n$. Moreover, in such an
application of the boundary method, a new series of intervals is generated, with
geometrically increasing lengths. These new intervals lie to the left, respectively
right, of $J_n^{(0)}$. They are now called $J_n^{(k)}$, with $k\le -1$ in the first case
and $k\ge 1$ in the second case. If an $I_n$ is modified twice, we also obtain two series
of intervals $J_n^{(k)}$, one for $k\le -1$ and another one for $k\ge 1$.

We define $\ell_n$ as the length of the final interval $J_n^{(0)}$. Then $J_n^{(0)}$ is
contained in the larger interval on which the estimates on $W$, $Q$ were originally
established -- compare \eqref{estWQ}, \eqref{1.12}; for example, $J_n^{(0)}\subset 3I_n$
if we were in case a) at that step. Since also $L_n\le \ell_n \le 2L_n$, we obtain the
following final estimates on the intervals $J_n^{(0)}$:
\[
\int_{J_n^{(0)}} W^2(x)\, dx \le \frac{268}{\ell_n},\quad
\int_{J_n^{(0)}} |Q(x)|\, dx \le \frac{562}{\ell_n} .
\]

Finally, we renumber everything so that $\ell_1\le \ell_2 \le \cdots$. In particular,
there are no gaps in this sequence now while in the original numbering, it can happen
that there are no $J_n^{(k)}$'s for a given $n\in\mathbb N$ (namely, if we were in case
b) or d) at step $n$).

These intervals $J_n^{(k)}$ certainly have the properties stated in part b) of
Theorem~\ref{TWQ}. Moreover, c) just records the estimates we have obtained above, with
all constants generously replaced by $10^3$. It is also clear that d) holds: Indeed,
$H_+$ or $H_-$ on $J_n^{(0)}$ has an eigenvalue $-E$ satisfying
\[
E\ge \frac{\epsilon_n}{2}=\frac{18}{L_n^2}
\ge \frac{18}{\ell_n^2} .
\]
By the min-max principle, the eigenvalues can only go up if Dirichlet boundary conditions
are introduced. Thus, since the $J_n^{(0)}$'s are disjoint and since the sequences
$\ell_n$ and $E_n^{-1/2}$ are non-decreasing, we must have that $\ell_n\ge 3\sqrt{2}
E_n^{-1/2}> 4E_n^{-1/2}$, as claimed.

It is clear from the construction that $W$ is absolutely continuous, and the remaining assertions
of part a) now follow by summing the bounds on $\int W^2$, $\int |Q|$. We of course use the now
established part d) here.
\end{proof}
We now discuss half-line problems with Dirichlet boundary conditions at the origin.
Later, in Section 6, we will also discuss a variant of the procedure we are about to
describe for Neumann boundary conditions in a related but simpler situation.

The additional issue that needs to be addressed here is the question of how an initial interval
$[0,L]$ should be handled.
We start with a tent $\varphi$ supported by $[0,3L_1]$ and equal to one on $[L_1,2L_1]$,
where $L_1=\epsilon_1^{-1/2}$. This gives the bounds
\begin{equation}
\label{2.15}
\int_{L_1}^{2L_1} W^2(x)\, dx \le \frac{4}{L_1},\quad
\int_{L_1}^{2L_1} |Q(x)|\, dx \le \frac{8}{L_1} .
\end{equation}
We now simply admit that we do not have such estimates on $[0,L_1]$, remove this interval
nevertheless, and continue as in the proof of Theorem~\ref{TWQ}. There is also no
guarantee that $H_+$ or $H_-$ will have a small eigenvalue on $[0,L_1]$, so we really
remove this interval without having achieved anything there. However, now that this has
been done, we are exactly in the situation from the proof of Theorem~\ref{TWQ}. If, at
some point, one of our intervals $\widetilde{I}_n$ lies close to $L_1$, we now can apply
the boundary method without any modifications because \eqref{2.15} gives us the required
a priori control in a neighborhood of the boundary point $L_1$.
\section{Pr\"ufer Variables}
Some of our results will depend on an analysis of the solutions to the Schr\"odinger
equation \eqref{se}. We are interested in positive energies $E$, and we write $E=k^2$,
with $k>0$. In order to study solutions $y(x,k)$ of \eqref{se}, we use the following
Pr\"ufer-type variables; these are particularly well adapted to the situation where
$V=W'+Q$. One may also say that we will not study \eqref{se} directly, but rather an
associated Dirac system. In any event, introduce the solution vector $Y$ as
\[
Y(x,k)= \begin{pmatrix} y(x,k) \\  (y'(x,k)-W(x)y(x,k))/k \end{pmatrix} ,
\]
and write
\[
Y(x,k) = R(x,k)
\begin{pmatrix} \sin (\psi(x,k)/2) \\ \cos (\psi(x,k)/2) \end{pmatrix} ,
\]
with $R(x,k) > 0$ and $\psi(x,k)$ continuous in $x$.
A computation shows that $R$, $\psi$ obey the following equations:
\begin{align}
\label{E:PRC} \left( \ln R(x,k)\right)' & =
-W(x)\cos\psi(x,k) + \frac{Q(x)-W^2(x)}{2k}\sin\psi(x,k) ,\\
\label{E:PTC} \psi'(x,k) & = 2k + 2W(x)\sin\psi(x,k) + \frac{Q(x)-W^2(x)}{k}(\cos\psi(x,k)-1) .
\end{align}
The last term in both equations is integrable if $\sum E_n^{1/2}<\infty$
and thus does not change the asymptotics.

For later use, we also note that $Y$ solves the first-order Dirac system
\[
Y'(x,k) = \left[ \begin{pmatrix} W(x) & k \\ -k & -W(x) \end{pmatrix}
+\frac{Q(x)-W^2(x)}{k} \begin{pmatrix} 0 & 0 \\ 1 & 0 \end{pmatrix} \right] Y(x,k) .
\]
Again, the $L_1$ term will be treated as a perturbation, and thus we will also study
the solutions $Y_0$ of the unperturbed system
\begin{equation}
\label{Y0}
Y'_0(x,k) = \begin{pmatrix} W(x) & k \\ -k & -W(x) \end{pmatrix} Y_0(x,k) .
\end{equation}
If we change the independent variable as follows,
\[
Y_0(x,k) = \begin{pmatrix} e^{ikx} & e^{-ikx} \\ ie^{ikx} & -ie^{-ikx} \end{pmatrix} Z(x,k),
\]
this becomes
\begin{equation}
\label{3.4}
Z'(x,k) = W(x) \begin{pmatrix} 0 & e^{-2ikx} \\ e^{2ikx} & 0 \end{pmatrix} Z(x,k) .
\end{equation}
\section{Proof of Theorem~\ref{Tess}}
Theorem~\ref{TWQ} shows that we can find disjoint intervals $I_n$ with lengths
$L_n\to\infty$, so that on $I_n$, we can write $V=W'+Q$ with $\int_{I_n} W^2,
\int_{I_n}|Q|\lesssim L_n^{-1}$. Actually, this only requires Lemmas~\ref{LWQ} and
\ref{Lremove} and not the full-fledged method from the proof of Theorem~\ref{TWQ}. The
easiest way to obtain such $I_n$'s is to work on the remaining half line at each step.

The Pr\"ufer equation \eqref{E:PTC} shows that
\[
\psi(a_n,k)-\psi(a_{n-1},k)= 2kL_n +
2\int_{a_{n-1}}^{a_n} W(x)\sin\psi(x,k)\, dx +O(L_n^{-1}) .
\]
Here $(a_{n-1},a_n)=I_n$ is one of the intervals from the preceding paragraph, and the constant
implicit in $O(L_n^{-1})$ remains bounded if $k>0$ stays away from zero.

Since $\|W\|_{L_1(I_n)} \le L_n^{1/2} \|W\|_{L_2(I_n)}\lesssim 1$, it follows that
\begin{equation}
\label{8.1}
\psi(a_n,k)-\psi(a_{n-1},k) = 2kL_n + O(1),
\end{equation}
with uniform control on the error term for $k\ge k_0>0$.

Since $y(x)=0$ precisely if $\psi(x)=2n\pi$ with $n\in\mathbb Z$ and thus
$\psi'(x)=2k>0$ at such a point, the Pr\"ufer angle $\psi$ may be used to
count the zeros of $y$.
Now suppose that $0<k_1<k_2$. Then \eqref{8.1} and the above remarks
show that for sufficiently large $n$,
every solution $y(\cdot,k_2)$ of the equation with $E=k_2^2$
has more zeros on $I_n$ than any non-trivial
solution $y(\cdot, k_1)$ for $E=k_1^2$. By Sturm comparison, $y(\cdot,k_2)$ then also has more zeros than
$y(\cdot,k_1)$ on $(0,x)$ for all large $x$.
Hence, by oscillation theory again, $[k_1,k_2]\cap\sigma\not=\emptyset$. This holds
whenever $0<k_1<k_2$, so $\sigma\supset [0,\infty)$.
\hfill$\Box$
\section{Spectral Properties}
In this section, we prove Theorems~\ref{TDe-K}, \ref{TC-K}, and \ref{Tdim}. The first two
proofs follow ideas of Deift-Killip \cite{DeK} and Christ-Kiselev \cite{CK,CK2},
respectively, rather closely. Our presentation will be very sketchy in these cases.

Let us begin with the \textit{proof of Theorem~\ref{TDe-K}.} Theorem~\ref{TWQ} shows that
$V=W'+W^2+Q-W^2\equiv W'+W^2+V_0$, with $V_0\in L_1$. This is a perturbation that is of
relative trace class in the form sense, so it suffices to consider the modified potential
$\widetilde{V}=W'+W^2$. Note that the form of $\widetilde{V}$ guarantees that
$-d^2/dx^2+\widetilde{V}$ on $L_2(\mathbb R)$ has no negative eigenvalues. Let us now
temporarily assume that $W$ is also of compact support $\subset (0,\infty)$. Then, as
$\int W'=0$, the first Faddeev-Zakharov trace formula \cite{ZF} reads
\begin{equation}
\label{ZFTF} \frac{1}{\pi} \int_{-\infty}^{\infty} \ln (1-|r(k)|^2)\, dk = -
\int_0^{\infty} W^2(x)\, dx .
\end{equation}
Here, the reflection coefficient $r$ is defined as $r=b/a$, where $f=ae^{ikx}+be^{-ikx}$
is the expansion of the solution $f$ close to zero that is equal to $e^{ikx}$ to the
right of the support of $W\!$. Now, given \eqref{ZFTF}, one can prove Theorem~\ref{TDe-K}
by following the arguments of \cite{DeK}: Approximate the actual $W$ in $L_2$ by
compactly supported $W_n$'s. The sum rule \eqref{ZFTF} provides a uniform lower bound on
the spectral densities of these problems. One then also needs weak $*$ convergence of the
spectral measures. We can again (as in \cite{DeK}) deduce this from the locally uniform
convergence of the $m$ functions on the upper half plane. This can most conveniently be
established in the framework of the associated Dirac system \eqref{Y0}. Since a compactly
supported perturbation does not change the absolutely continuous spectrum, we may also
assume that $0\notin\text{supp }W\!$. This is helpful here because then
$Y(0)=(y(0),y'(0)/k)^t$, and hence the $m$ functions of the original Schr\"odinger
equation and the associated Dirac system are directly related. We conclude our sketch of
the proof with these remarks. The reader may also wish to consult the introduction of
\cite{MTT2} for further background information.\hfill$\Box$

\medskip

We now want to analyze the situation under the stronger hypotheses of Theorems~\ref{TC-K}
and \ref{Tdim}. The notation from Theorem~\ref{TWQ} is too clumsy for this purpose. So we
reorganize the intervals $J_n^{(k)}$ once more and now simply denote them by
$I_n=(a_{n-1},a_n)$, with $0=a_0<a_1<\cdots$. We also write $L_n=|I_n|$. So
Theorem~\ref{TWQ}c) (or rather the half-line version discussed at the end of Section 2)
now says that
\begin{equation}
\label{5.1}
\int_{I_n} W^2 \lesssim L_n^{-1}, \quad\quad \int_{I_n} |Q| \lesssim L_n^{-1} \quad (n\ge 2) .
\end{equation}

The following observation is essential for what follows: If $\sum E_n^p<\infty$ ($p>0$),
then also $\sum L_n^{-2p}<\infty$. To prove this, note that by their definition, the
lengths $L_n$ are a rearrangement of $|J_n^{(k)}|$. Moreover, Theorem~\ref{TWQ}b), d)
shows that $|J_n^{(k)}|\gtrsim 2^{|k|} E_n^{-1/2}$. Thus
\[
\sum_{n=1}^{\infty} L_n^{-2p} \lesssim \sum_{n=1}^{\infty} \sum_{k=-N_n}^{N'_n} 2^{-2p|k|}E_n^p ,
\]
and our claim follows.

\medskip

Now, given Theorem~\ref{TWQ}, the \textit{proof of Theorem~\ref{TC-K}} consists of not
much more than a quotation: Assume that $\sum E_n^p<\infty$ with $0<p<1/2$. H\"older's
inequality shows that
\[
\int_{I_n} |W|^{2p+1} \le \left( \int_{I_n} W^2 \right)^{p+1/2} L_n^{1/2-p}
\lesssim L_n^{-2p} .
\]
Thus $W\in L_q$ for some $q<2$. Now the machinery of Christ-Kiselev \cite{CK,CK2} gives
the desired asymptotics for the solutions $Y_0$ of the unperturbed system \eqref{Y0}; see
especially the system (2.2) from \cite{CK} and the analysis that follows. Here, we
instead use \eqref{3.4} as our starting point. Actually, the situation is simpler than in
\cite{CK} because we have Fourier transforms instead of WKB transforms. To also obtain
the corresponding asymptotics for $Y\!$, one can finally use a standard perturbative
argument based on Levinson's Theorem \cite[Theorem 1.3.1]{East} (compare again
\cite{CK}).\hfill$\Box$

\begin{proof}[Proof of Theorem~\ref{Tdim}.]
We now assume that $\sum E_n^p<\infty$, with $0<p<1/4$. As observed above, this implies that
$\sum L_n^{-2p}<\infty$.
We will use ideas
from \cite{dk,Remac,Remdim}. Also, the following estimate will be a crucial ingredient.
\begin{Lemma}
\label{Lcap} Let $\alpha\in (0,1)$. Then there exists a constant $C_{\alpha}$ so that for
all $f\in L_2(a,b)$, all finite Borel measures $\mu$ on $\mathbb R$, and all measurable
functions $c$ with $a\le c(k) \le b$, the following estimate holds $(L\equiv b-a)$:
\[
\int d\mu(k)\, \left| \int_a^{c(k)} dx\, f(x) e^{2ikx} \right| \le C_{\alpha}
\mathcal{E}_{\alpha}^{1/2}(\mu) L^{(1-\alpha)/2} \|f\|_{L_2(a,b)} .
\]
Here $\mathcal{E}_{\alpha}(\mu)\equiv \int\!\!\int d\mu(k)\, d\mu(l) (1+|k-l|^{-\alpha})$
denotes the $\alpha$ energy of $\mu$.
\end{Lemma}
This follows by slightly adjusting the calculation from \cite[vol.~II, pg.~196]{Zyg},
so we will not give the proof here. Compare also \cite[Lemma 7.6]{dk}.

Note that if for some $E>0$, we have two linearly independent solutions for which the limits
$\lim_{x\to\infty} R(x,k)$, $\lim_{x\to\infty} (\psi(x,k)-2kx)$ exist, then it follows
(by taking a suitable linear combination) that $E\notin S$. Thus it suffices to show that
for arbitrary but fixed initial
values $R(0,k)$, $\psi(0,k)$, these limits exist off a set of dimension
at most $4p$.
We will split the proof of this into two parts. This is not really necessary, but
it will help to make the presentation more transparent.
In the first step, we will show that the limits exist on the subsequence $a_n$ as long as we
stay off an exceptional set. In the second step, we will extend this to sequences tending to
infinity arbitrarily. Actually, rather similar arguments are applied in both steps, so the
second step will not be very difficult once we have completed the first step.

So in this first step, we are concerned with the series
\begin{equation}
\label{4.1}
\sum_{n=1}^{\infty} \left| \int_{a_{n-1}}^{a_n} W(x) e^{i\psi(x,k)}\, dx \right| .
\end{equation}
Indeed, real and imaginary parts of the integrals give us the leading terms from the
equations for $R$ and $\psi-2kx$, respectively; see \eqref{E:PRC}, \eqref{E:PTC}. So it
suffices to show that this series converges off an exceptional set of dimension $\le 4p$.

We will need control on the maximal function
\[
M_n(k) \equiv \max_{a_{n-1}\le c\le a_n} \left| \int_{a_{n-1}}^c W(x)e^{2ikx}\, dx \right| .
\]
Let $\mu$ be any (Borel) measure with finite $4p$ energy. Since $\|W\|_{L_2(a_{n-1},a_n)}
\lesssim L_n^{-1/2}$ by \eqref{5.1}, Lemma~\ref{Lcap} says that $\|M_n\|_{L_1(\mu)}
\lesssim L_n^{-2p}$. Now $\sum L_n^{-2p}<\infty$, so the Monotone Convergence Theorem
shows that $M_n(k)\in\ell_1$ for $\mu$ almost every $k$. Since $\mu$ is only assumed to
have finite $4p$ energy but is otherwise arbitrary, it follows that $\dim S_0\le 4p$,
where
\[
S_0 = \left\{ k>0: \sum_{n=1}^{\infty}  M_n(k)= \infty \right\} .
\]
This conclusion is nothing but a standard relation between capacities and Hausdorff dimensions; for
example, one can argue as follows: Suppose that, contrary to our claim, $\dim S_0 > 4p$, and fix
$d\in (4p,\dim S_0)$. Then,
since $S_0$ is a Borel set of infinite $d$-dimensional Hausdorff measure, there exists a finite measure
$\mu\not= 0$ supported by $S_0$ with $\mu(I) \le C|I|^d$ for all intervals $I\subset\mathbb R$
\cite[Theorem 5.6]{Falc}. It is easily seen that $\mathcal{E}_{4p}(\mu) <\infty$ for such a $\mu$.
So what we have shown above now says that $M_n(k)\in\ell_1$ for $\mu$ almost every $k$, which clearly
contradicts the fact that $\mu$ is supported by $S_0$.

We now claim, more specifically, that \eqref{4.1} converges if $k\notin S_0$.
Write $\psi = 2kx +\varphi$, and consider one of the integrals
$\int_{a_{n-1}}^{a_n} W(x)e^{i\psi(x,k)}\, dx$ from \eqref{4.1}.
Integration by parts gives
\begin{multline*}
\int_{a_{n-1}}^{a_n} W(x)e^{i\psi(x,k)}\, dx = e^{i\varphi(a_n,k)}\int_{a_{n-1}}^{a_n} W(x)e^{2ikx}\, dx \\
- i \int_{a_{n-1}}^{a_n} dx\, e^{i\varphi(x,k)}(2W(x)\sin\psi(x,k)+\rho(x))
\int_{a_{n-1}}^x dt\, W(t) e^{2ikt} .
\end{multline*}
We have abbreviated the integrable term as $\rho=(1/k)(Q-W^2)(\cos\psi-1)$.
The first term on the right-hand side is summable for $k\notin S_0$, so
we must now show that the series over the second term on the right-hand side
is also absolutely convergent for these $k$.
This, however, is immediate from the bound
\[
\left( 2\|W\|_{L_1(a_{n-1},a_n)} + \|\rho\|_{L_1(a_{n-1},a_n)} \right) M_n(k)
\]
on this term since $\|W\|_{L_1(a_{n-1},a_n)}\lesssim 1$.

So we know now that $\lim_{n\to\infty} R(a_n,k)$, $\lim_{n\to\infty} (\psi(a_n,k)-2ka_n)$
exist for all $k\notin S_0$. As the second step of the proof of Theorem~\ref{Tdim}, we
need to extend this to sequences tending to infinity in an arbitrary way. This suggests
that we look at
\[
\widetilde{M}_n(k)\equiv \max_{a_{n-1} \le c\le a_n} \left| \int_{a_{n-1}}^c W(x)e^{i\psi(x,k)}\, dx
\right| .
\]
We claim that $\widetilde{M}_n(k)\to 0$ if $k\notin S_0$. This will complete the proof of
Theorem~\ref{Tdim} because it will then follow that $S\subset S_0$.

To prove our claim on $\widetilde{M}_n(k)$, we proceed as above and reduce matters to
the corresponding statement on $M_n(k)$, which we know is true. This does not require new ideas.
We just repeat the above computations, but with the upper limit $a_n$ now replaced by
$c=c(k)$. Everything goes through as before, and we have in fact proved the
stronger statement that $\widetilde{M}_n(k)\in \ell_1$ if $k\notin S_0$.\hfill$\Box$

\section{Sign-Definite Potentials}
We prove Theorem~\ref{TILT} here. We use the strategy from Section~2. The treatment
simplifies considerably because there is no need to resort to Lemma~\ref{LWQ}. Indeed, if
$H_+\ge -\epsilon$ on $I$, then
\begin{equation}
\label{6.1} -\int V\varphi^2 \le \epsilon\int\varphi^2 + \int\varphi'^2
\end{equation}
for all test functions $\varphi\in H_1(I)$ that vanish at the finite endpoints of $I$.
Since $V\le 0$ now, this may be used to bound the $L_1$ norm of $V$ over suitable
intervals. Therefore, the proof of Theorem~\ref{TWQ} now produces intervals $J_n^{(k)}$
with the same geometry as before, and $\int_J |V|\lesssim 1/|J|$. In particular, if $\sum
E_n^p<\infty$, then also $\sum |J_n^{(k)}|^{-2p}<\infty$, and this latter sum may be
estimated by a multiple of the first sum. The constant only depends on $p$. This follows
as usual from $|J_n^{(k)}|\gtrsim 2^{|k|}E_n^{-1/2}$ by first summing over $k$ and then
over $n$.

Theorem~\ref{TILT}a) for the whole-line problem is now immediate from H\"older's
inequality, which gives, for $0<p\le 1/2$,
\[
\int_J |V|^{p+1/2} \le \left( \int_J |V| \right)^{p+1/2} |J|^{1/2-p}
\lesssim |J|^{-2p} .
\]
The proof of part b) is similar. Since $|J_n^{(k)}|\gtrsim E_1^{-1/2}\ge E_0^{-1/2}$,
we have that for $p\ge 1/2$,
\[
\sum \left( \int_n^{n+1} |V| \right)^{2p} \lesssim \sum_{n,k} \left( \int_{J_n^{(k)}} |V|
\right)^{2p}
\]
with a constant that depends on $E_0$.

To prove Theorem~\ref{TILT} on the half line with Neumann boundary conditions at the
origin, we use a variant of the argument from the end of Sect.~2. Namely, let again
$L_1=\epsilon_1^{-1/2}$. However, due to the Neumann boundary conditions (instead of
Dirichlet), we can use a test function $\varphi$ now defined as follows: $\varphi=1$ on
$(0,2L_1)$, $\varphi=0$ on $(3L_1,\infty)$, and $\varphi$ is linear on the remaining
piece. Then \eqref{6.1} yields
\[
\int_0^{2L_1} |V(x)|\, dx \le \frac{4}{L_1} .
\]
As discussed at the end of Sect.~2, we can remove $(0,L_1)$ and run the algorithm from
the proof of Theorem~\ref{TWQ} on the remaining half line. (In particular, we impose
\textit{Dirichlet} boundary conditions at $x=L_1$.) In contrast to the situation there,
we now have control on $V$ on $(0,L_1)$ as well. So Theorem~\ref{TILT} for the half line
with Neumann boundary conditions now follows as above.
\end{proof}
\section{Counterexamples}
In this section, we prove Theorems~\ref{Tdimopt} and \ref{TL1opt}. In both cases, we use
sparse potentials and rely on previous work on the spectral properties of these models
\cite{kls,Remtams}. We then need control on the discrete spectrum, but, fortunately, this
is easy. We consider two different types of bumps. Let
\[
V_g(x) = -g\chi_{(-1,1)}(x),\quad W_g(x)= g \left( \chi_{(-1,0)}(x)-\chi_{(0,1)}(x) \right) .
\]
\begin{Lemma}
\label{Lsparse} {\rm a)} For small $g>0$, the operator $-d^2/dx^2 + V_g(x)$ on
$L_2(\mathbb R)$ has precisely one eigenvalue $-E$, and $E=E(g)=g^2+O(g^3)$.

{\rm b)} For small $g>0$, the operator $-d^2/dx^2 + W_g(x)$ on $L_2(\mathbb R)$ has
precisely one eigenvalue $-E$, and $E=E(g)=g^4/9+O(g^5)$.
\end{Lemma}
\begin{proof}[Sketch of the proof.]
It is clear that $-d^2/dx^2+V_g(x)$ has at most one eigenvalue for small $g>0$. This
follows from an elementary analysis of the solutions at zero energy (the number of zeros
on $(-N,\infty)$ of a solution $y$ with $y(-N)=0$ is the number of negative eigenvalues
of the operator on that interval). Since $W_g\ge V_g$, $-d^2/dx^2+W_g(x)$ has at most one
eigenvalue for small $g$. That the operators have at least one eigenvalue for $g>0$ is a
classical result for sign-definite potentials \cite[Theorem XIII.11]{RS4} and follows
from recent work \cite{dhks,dks} (and, incidentally, also from our Lemma~\ref{LWQ}) for
arbitrary potentials.

To approximately compute this eigenvalue $-E$, we note that the corresponding eigenfunction
$y$ must be a multiple of $e^{-E^{1/2}|x|}$ for $x\le -1$ and $x\ge 1$; the constant factors
may be different on these two half lines. So $-E$ is an eigenvalue precisely if there
exists $c\in\mathbb R$ so that
\begin{equation}
\label{9.1}
c \begin{pmatrix} 1 \\ -E^{1/2} \end{pmatrix} = T_g(1,-1;-E) \begin{pmatrix}
1 \\ E^{1/2} \end{pmatrix} .
\end{equation}
Here, $T_g$ is the transfer matrix, that is, the matrix that takes solution vectors
$(y,y')^t$ at $x=-1$ to their value at $x=1$. We of course have explicit formulae
for the transfer matrices for $V_g$ and
$W_g$, respectively, and a somewhat cumbersome but completely elementary discussion
of \eqref{9.1} then establishes the asserted asymptotics of $E$.
\end{proof}
Now consider a (half-line) potential $V$ of the form
\begin{equation}
\label{9.2}
V(x) = \sum_{n=1}^{\infty} V_{g_n}(x-x_n),
\end{equation}
with $g_n\to 0$. The $x_n$'s are typically very rapidly increasing so that the individual
bumps are well separated and thus almost independent of one another. To rigorously
analyze $V\!$, we build it up successively. The following lemma describes this situation.
\begin{Lemma}
\label{L9.2} Consider $H_a=-d^2/dx^2+Q(x)+V_g(x-a)$, where $Q$ has compact support and
$g$ is sufficiently small so that Lemma~\ref{Lsparse} applies. Suppose that
$-d^2/dx^2+Q(x)$ has precisely $N$ negative eigenvalues
$-\widetilde{E}_1,\ldots,-\widetilde{E}_N$ on $L_2(0,\infty)$. Then, for any
$\epsilon>0$, there exists $a_0$ so that for all $a\ge a_0$, the following holds: $H_a$
on $L_2(0,\infty)$ has precisely $N+1$ negative eigenvalues $-E_1,\ldots,-E_{N+1}$, and
\[
\left| E_i -\widetilde{E}_i\right|<\epsilon,\quad \left| E_{N+1} - E(g)\right|
<\epsilon .
\]
An analogous statement holds for $H_a$ on $L_2(0,2a)$.
\end{Lemma}
In other words, there is almost no interaction between $Q$ and $V_g$ and the
eigenvalues approximately behave like those of
the orthogonal sum of $-d^2/dx^2+Q(x)$ and $-d^2/dx^2+V_g(x)$.
\begin{proof}
Let $y$ be the solution of $-y''+(Q+V_g)y=0$ with $y(0)=0$, $y'(0)=1$. By taking $a$
large enough, we can make sure that $y$ has precisely $N$ zeros on $(0,a/2)$ (say) and
$y(a/2)y'(a/2)\ge 0$, $y(a/2)\not= 0$ (if $y$, $y'$ had different signs, there would be
another zero on $(a/2,\infty)$ for zero potential). Here we again use oscillation theory;
more precisely, the number of positive zeros of $y$ is exactly the number of negative
eigenvalues. An elementary discussion now shows that $y$ has exactly one more zero on
$(a/2,\infty)$ if $a$ is large enough. Thus $H_a$ has precisely $N+1$ negative
eigenvalues. Since this additional zero in fact lies in $(a+1,2a)$, this also holds for
$H_a$ on $L_2(0,2a)$.

It is now easy to approximately locate these eigenvalues: Cut off the eigenfunctions of
$-d^2/dx^2+Q(x)$ (on $L_2(0,\infty)$) and of $-d^2/dx^2+V_g(x-a)$ (on $L_2(\mathbb R)$)
by multiplying by suitable smooth functions which are equal to one except in a
neighborhood of, say, $a/2$ (and $2a$ for this latter eigenfunction). If $a$ is large
enough, we thus obtain functions $\varphi_i$ with
\[
\|(H_a+\widetilde{E}_i)\varphi_i\|
<\epsilon\|\varphi_i\| \quad (i=1,\ldots,N),\quad\quad \|(H_a+E(g))\varphi_{N+1}\|<\epsilon
\|\varphi_{N+1}\| .
\]
This completes the proof of the lemma.
\end{proof}

We now claim that given any sequence $\epsilon_n>0$, we can find $x_n$'s in \eqref{9.2},
so that $-d^2/dx^2+V(x)$ has eigenvalues $-E_n$, satisfying $|E_n-E(g_n)|<\epsilon_n$,
and no other negative eigenvalues. To prove this, we may assume that $\epsilon_n$
decreases. To simplify the book keeping, we in fact further assume that the intervals
$(E(g_n)-\epsilon_n,E(g_n)+\epsilon_n)$ are disjoint. This is to say, we assume that the
$g_n$'s are distinct and then further decrease the $\epsilon_n$'s if necessary.

Now apply Lemma~\ref{L9.2} with $Q=0$, $g=g_1$, and $\epsilon= 2^{-2}\epsilon_1$ to find
$x_1$. In the second step, apply the lemma with $Q=V_{g_1}(x-x_1)$, $g=g_2$, and
$\epsilon=2^{-3}\epsilon_2$ (which is also $\le 2^{-3}\epsilon_1$ by our assumption).
This gives $x_2$; we may also demand that $x_2-1>2x_1$. Continue in this way. The
construction ensures that $-d^2/dx^2+V(x)$ on $L_2(0,2x_n)$ has precisely $n$ negative
eigenvalues $E_i^{(n)}$, and these satisfy $|E_i^{(n)}-E(g_i)|\le\epsilon_i/2$. Moreover,
for fixed $i$, $E_i^{(n)}$ is a Cauchy sequence and hence convergent. Since every
eigenvalue of the half-line problem is an accumulation point of eigenvalues of the
problems on $(0,2x_n)$, only the limits $E_i\equiv\lim_{n\to\infty} E_i^{(n)}$ can be
eigenvalues of $-d^2/dx^2+V(x)$. On the other hand, the half-line problem has at least as
many eigenvalues $<-E$ as the corresponding problem on $(0,2x_n)$; thus, since the
$E_i$'s are from the disjoint intervals $(E(g_i)-\epsilon_i,E(g_i)+\epsilon_i)$ and the
problems on $(0,2x_n)$ have spectrum in these intervals for large $n$ by construction and
Lemma~\ref{L9.2}, the negative spectrum is precisely the set $\{ E_i\}$.

Let us now prove Theorem~\ref{TL1opt}. Given $e_n>0$ with $e_n\to 0$, $\sum e_n=\infty$,
pick $g_n>0$ so that $E(g_n)=e_n/2$, say. By slightly decreasing the $e_n$'s if
necessary, we may of course assume that the $e_n$'s and hence also the $g_n$'s are
distinct. By Lemma~\ref{Lsparse}a), the $g_n$'s will satisfy $g_n\sim (e_n/2)^{1/2}$ in
the sense that the ratio tends to one. Now choose $x_n$'s as above so that the operator
$-d^2/dx^2+V(x)$ with $V$ as in \eqref{9.2} has eigenvalues $E_n \le e_n$. Since $V\le
0$, $H_-=-d^2/dx^2-V(x)$ has no negative spectrum. When choosing the $x_n$'s, we can
further require that $x_n/x_{n+1}\to 0$. Since also $\sum g_n^2=\infty$, a well-known
result on sparse potentials applies and the spectrum is purely singular continuous on
$(0,\infty)$ (see \cite[Theorem 1.6(2)]{kls}).

The proof of Theorem~\ref{Tdimopt} uses $W_g$ and \cite[Theorem 4.2b)]{Remtams} instead
of $V_g$ and \cite[Theorem 1.6(2)]{kls}, respectively, but is otherwise analogous. First
of all, the analog of Lemma~\ref{L9.2} holds, with a similar proof. Let $e_n>0$ with
$\sum e_n^{1/4}=\infty$ be given. If $e_n$ is non-increasing, we also have that $\sum
e_{2n}^{1/4}=\infty$. Determine $g_n$'s so that $E(g_n)=e_{2n}/9$, where $E(g)$ now
refers to $W_g$. Lemma~\ref{Lsparse}b) shows that then $g_n\sim e_{2n}^{1/4}$. Define $V$
as in \eqref{9.2}, but with $W_g$ instead of $V_g$. We can then find $x_n$'s so that
$-d^2/dx^2+V(x)$ has eigenvalues $E_n\le e_{2n}$ in this case as well, following the same
arguments as above. Moreover, since $-d^2/dx^2-W_g(x)$ on $L_2(\mathbb R)$ has the same
eigenvalue $-E(g)$ as $-d^2/dx^2+W_g$, we can also arrange that $-d^2/dx^2-V(x)$ has
eigenvalues $E'_n \le e_{2n}$. Thus, after combining the eigenvalues of $H_{\pm}$ in one
sequence $E_n$, we have that $E_n\le e_n$, as desired. Finally, we can again require that
$x_n/x_{n+1}\to 0$. Since $\sum g_n=\infty$, Theorem 4.2b) from \cite{Remtams} applies
and shows that $\dim S=1$.

\end{document}